\documentclass{article}
\usepackage[utf8]{inputenc}
\usepackage{hyperref}
\usepackage{blindtext}
\usepackage{titlesec}
\usepackage{setspace}
\usepackage{float}
\usepackage{graphicx}
\usepackage[margin=1.2in]{geometry}
\usepackage[labelfont=bf]{caption}
\usepackage{amsmath}
\usepackage{amsfonts}
\usepackage{amssymb}
\usepackage{indentfirst}
\usepackage{verbatim}
\usepackage{afterpage}

\immediate\write18{texcount -tex -sum  \jobname.tex > \jobname.wordcount.tex}

\providecommand{\keywords}[1]
{
  \small	
  \textbf{\textit{Keywords---}} #1
}

\title{\textbf{Convolutional encoder decoder network for the removal of coherent seismic noise}}
\author{\textbf{Yash Agarwal$^{1}$, Sarah Greer$^{2}$}  \\
        \small $^{1}$Dougherty Valley High School, San Ramon, CA 94582 \\
        \small $^{2}$Massachussets Institute of Technology, Cambridge, MA, 02139 \\
}
\date{} 

\begin{document}
\newpage
\maketitle


\begin{abstract}
  Seismologists often need to gather information about the subsurface structure of a location to determine if it is fit to be drilled for oil. However, there may be electrical noise in seismic data which is often removed by disregarding certain portions of the data with the use of a notch filter. Instead, we use a convolutional encoder decoder network to remove such noise by training the network to take the noisy shot record as input and remove the noise from the shot record as output. In this way, we retain important information about the data collected while still removing coherent noise in seismic data.
\end{abstract}
\tableofcontents
\newpage
\section{Introduction}
\setlength{\parindent}{5ex}

Seismic experiments are often used in oil and gas exploration to look for subsurface oil deposits. 
In a typical seismic experiment, sources and receivers are placed on the surface over a domain of interest. 
A wave propagates from a source location, interacts with the underlying discontinuities in the subsurface, and arrives back to the surface to be recorded by the receivers.
These data are typically used to produce an image of the subsurface, which aims to show the underground geologic structure below the area of interest.

These datasets are often corrupted with ambient random background noise, but can also have coherent noise. 
The added noise can be difficult to remove, and can diminish the relative power of the signal of interest which contains the desired information about the subsurface structures. 
In this paper, we look at the problem of removing powerline, or electrical, noise from seismic data.
When a receiver is placed near an electrical device which uses an alternating current, it oftentimes picks up a characteristic 60 Hz (or 50 Hz, depending on locality) sinusoid element.
This sinusoid can cover up the signal from the seismic experiment, which contains the desired information about the subsurface structure.
As such, it is best practice to remove this noise before further processing is done.
Traditionally, since the noise has a narrow bandwidth, this is done by applying a notch filter where all the signal around 60 Hz is decimated.
While this can successfully remove the powerline noise, it also removes valuable information from the signal of interest which is contained within the same bandwidth that is removed.

Our goal is to find a way of removing the characteristic 60 Hz powerline noise while still retaining the signal information at the 60 Hz band. 
Previous methods have looked at using various filters \cite{mean, hum}, subtracting estimates of the noise from the data \cite{subtract}, and using randomized principal component analysis \cite{rpca}.
We take a different approach, and aim to do this using a convolutional encoder decoder network.

\section{Data Generation}
\begin{figure}[H]
  \centering
  \includegraphics[width=0.74\textwidth]{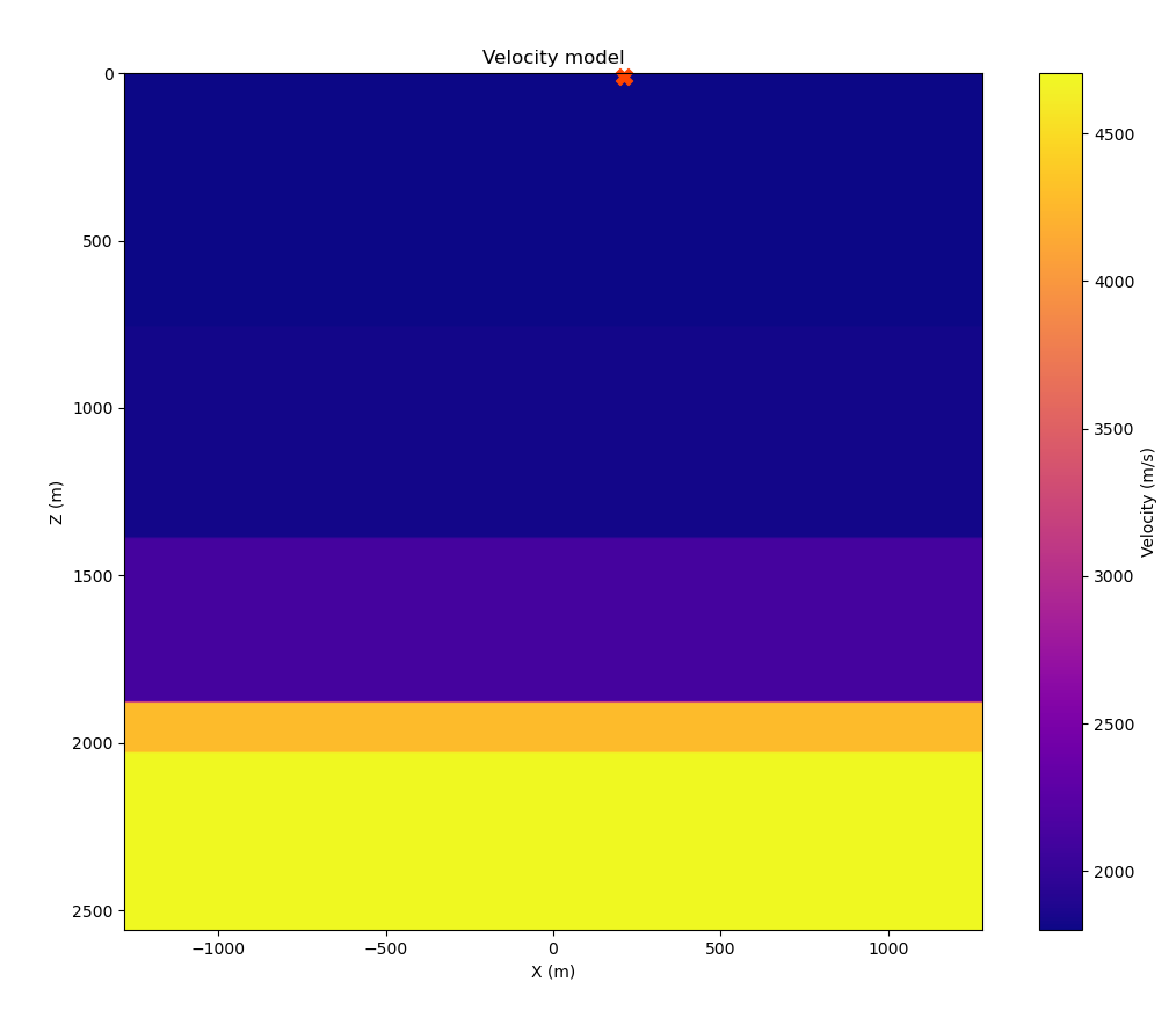}
  \caption{An example velocity model from our generated dataset. The colorbar on right depicts the velocity (in $m/s$) of P-wave propagation for the respective layer. The X at the top depicts an example location of a source at $250$ meters, from which a single shot record is created.}
  \label{fig:testfig}
\end{figure}

\begin{figure}[H]
  \centering
  \includegraphics[width=\textwidth]{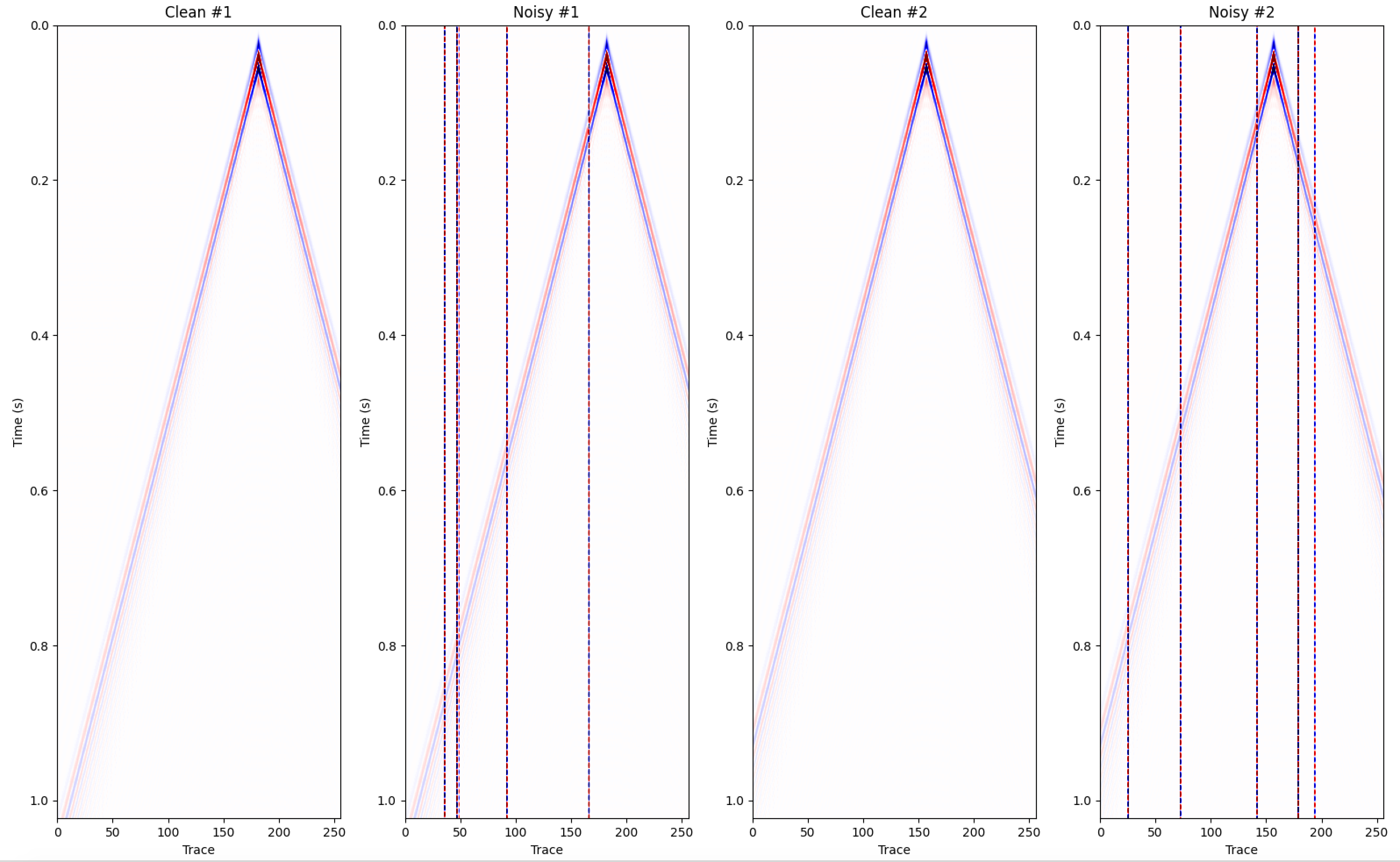}
  \caption{Clean \#1 and Clean \#2 show two random images from the RegShot dataset. The corresponding shot records with added noise are images Noisy \#1 and Noisy \#2, which are part of the NoisyShot dataset. The coherent electrical noise at 60 Hz is present in the form of vertical dashed lines.}
  \label{fig:sroriginal}
\end{figure}

We create a dataset of flat, or ``layer-cake'', velocity models, which consists of $2000$ randomly generated $1024\times256$ images with layer velocities ranging from $1800$  $m/s$ to $5000$ $m/s$ and thickness ranging from $50$ meters to $400$ meters. The velocity values of the layers are sorted from least to greatest from top to bottom to account for the fact that subsurface velocity typically increases with depth. We put sources (shots) along the surface and spaced approximately $250$ meters apart (specifically, we put sources at $-500, -250, 0, 250,$ and $500$ meters). For every velocity model generated, we run 5 different experiments and thus produce 5 different shot records---one for each source. We placed receivers across the surface $10$ meters apart. These velocity models represent the velocity of the P-wave propagated through the subsurface at different locations of a 2D cross section of the subsurface. Note that these velocity models are in a layer-like format because of the fact that sediment is usually deposited as layers, and the speed of a P-wave has a different velocity when traveling through different types of rocks. An example of a velocity model is shown in Figure \ref{fig:testfig}. 
From the 2,000 velocity models and 5 shots per velocity model, we create a dataset of $10,000$ corresponding shot records, which we labeled RegShot. We then generate another dataset incorporating powerline noise randomly for all of the data in RegShot, which we labelled NoisyShot. The noise is added by randomly selecting between 3 and 6 traces for each shot gather, and adding a unique sinusoid $A \sin(2\pi\, f\; t + \phi)$ to each trace. 
Here, $A$ is randomly scaled to be within normal range based on the amplitude of the data, $f$ is 60 Hz, $t$ is time, and the phase $\phi$ is randomly chosen to be between $0$ and $2 \pi$. 
Our forward modeling, which generates our RegShot dataset with the input of a given velocity model and survey geometry, is generated using a fourth order finite difference scheme. 
We use a 25 Hz Ricker wavelet as our source, which is a standard  wavelet used in exploration seismology.
We use the data from NoisyShot as our input to the network, and the corresponding noise-free shot gather from RegShot is our desired truth image.
Examples of images from the RegShot and NoisyShot datasets are pictured in Figure \ref{fig:sroriginal}.
The horizontal axis of the shot record is where the $256$ receivers lie, and all $256$ of their recordings of the P-wave together create the shot record. The vertical axis is the time in seconds. 
\section{Our Method}
\subsection{Pre-processing}
We normalized our data using \emph{z}-score normalization \cite{zscore}. We found the mean and standard deviation of the distribution of all pixel values for all $10,000$ shot records in the NoisyShot dataset and changed all pixel values as
\[\emph{z} = \frac{x-\mu}{\sigma} \]
where $\mu$ is the population mean of all the pixels of all the shot records in the NoisyShot dataset, $\sigma$ is the population standard deviation of all of these pixel values, $x$ is the original pixel value, and $z$ represents the new pixel value after \emph{z}-score normalization. We found the mean pixel value to be approximately $7.55253202*10^{-7}$ and the standard deviation of the pixel values to be approximately $2.68891895$.
\subsection{Network Structure}
\begin{figure}[H]
  \centering
  \includegraphics[width=0.96\textwidth]{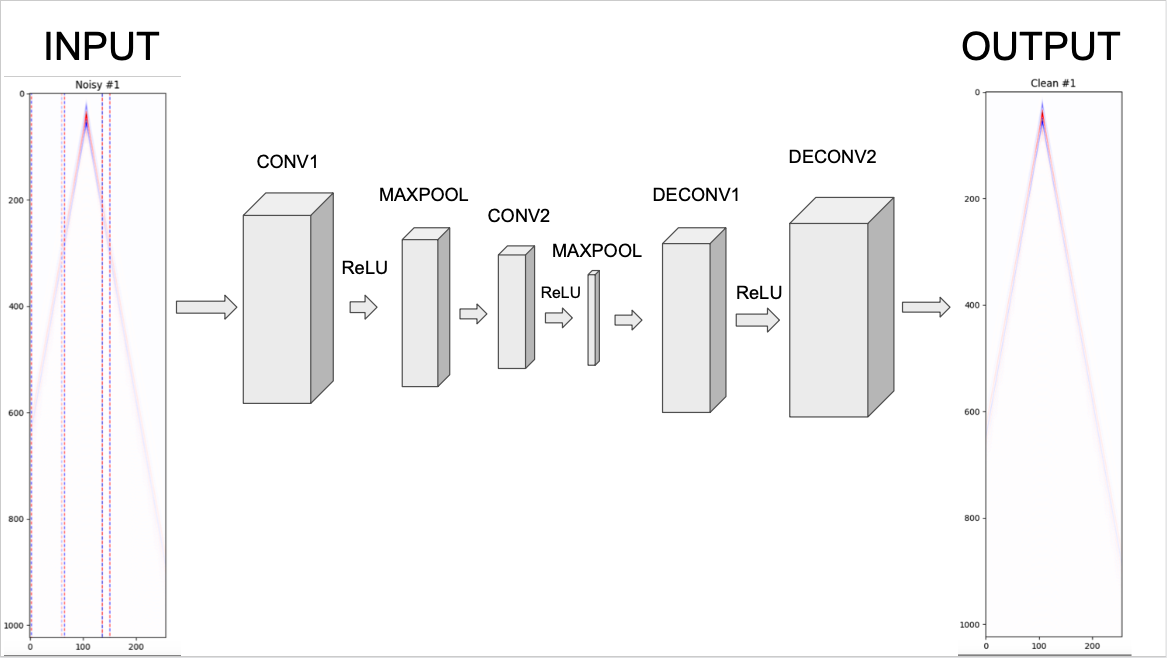}
  \caption{The figure depicts the structure of our encoder-decoder network. We use two convolutional layers and two deconvolutional layers with Max Pooling layers in between. The goal of the encoder-decoder network is to take in the noisy shot gather as input and to output the clean shot gather.}
  \label{fig:structure}
\end{figure}

We used a convolutional encoder-decoder network to remove such coherent noise in the noisy shot records, as the convolution operation is translation invariant, which means that the detection of features in the shot record will not be affected by the position of the seismic wave. This is critical as we are using $5$ sources for every velocity model, which results in 5 shot records which, in the case of solely horizontal layers, are translations of one another. 
The network structure is summarized in Figure \ref{fig:structure}.
We use two convolutional layers in the encoder and two deconvolutional layers in the decoder, as well as pooling layers and activation functions in between. 
\subsubsection{Encoder}
First, we describe the encoder portion of our network, also known as the process of downsampling. The goal of the encoder is to reduce the dimensionality of the seismic data while trying to remove the coherent noise present in the data. The convolutional layers use convolution kernels to try and understand patterns in the data (features). A feature map is then created as the kernel convolves with the input data. The equation of the convolution operation for the two convolutional layers is
\[F_k(x_{i,j}^l) = w_k^lx_{i,j}^l+b_k^l\;,\]
where we define a function $F_k: \mathbb R^2 \to \mathbb R^2$ which represents the value of the feature at the point $(i,j)$ of the $k$th feature map of the $l$th layer, $w_k^l$ is the kernel of the $k$th filter of the $l$th layer, and $b_k^l$ is the bias term of the $k$th filter of the $l$th layer of our encoder network \cite{math}. We then introduce nonlinearity by using the ReLU activation function, which is given by the following piecewise function:
\begin{equation}
\text{ReLU}(x) =   \left\{
\begin{array}{ll}
      0 & x < 0 \\
      x & x \geq 0 \\
\end{array}\;. 
\right.
\end{equation}
Finally, we use a max pooling layer, where we divide the input image into different regions (based on the size of our kernel) and we take the maximum feature value of each region. The two convolutional layers in the encoder have a $3 \times 3$ kernel, with a padding of $1$ and a stride of $1$. We then use a max pooling layer, which has a kernel of dimension $2 \times 2$ and a stride of $2$.
\subsubsection{Decoder}
We describe the decoder portion of our network, also known as the process of upsampling. The goal of the decoder is to produce an image as close to the original image as possible without the added coherent noise. We use two deconvolutional layers as part of the decoder of our network, with the ReLU activation function in between. The two deconvolutional layers use transposed convolutions, which uses the transpose of the sparse matrix which is represented by the output of the convolutional layers. In this way, they preserve the connectivity pattern of the convolution and thus create as close of an image to the original as possible \cite{arithmetic}. These layers have a kernel size of $2 \times 2$ and a stride of $2$.
\subsection{Objective Function/Optimizer}
We used a mean square error (MSE) loss function as our data largely consists of $0$ values while having very large and very small values as part of the seismic wave propagation. Thus, since MSE loss penalizes outliers more than other loss functions such as mean absolute error (MAE), we chose to use MSE. Since the corrupting electrical noise is frequency dependent, analyzing the performance of the model in the frequency domain can provide a more complete picture of model performance. Because of this, we consider the performance of the model in the time and frequency domains. Our objective function is:
\begin{equation*}
\L(\hat{y_i}, y_i) =\sum_{i=1}^{N} \left [\lVert\hat{y_i}-y_i\rVert_{2}^{2} + \lVert\mathcal{F}(\hat{y_i})-\mathcal{F}(y_i)\rVert_{2}^{2} \right ] \;,
\end{equation*}
where we sum over all $N$ shot records in each of the $32$ batches for the training and test data, $y_i$ is the set of pixel values from an image of the RegShot dataset (truth image), $\hat{y_i}$ is the set of pixel values from an image predicted by our network from an image of the NoisyShot dataset, and $\mathcal{F}$ is the Fourier Transform function. In other words, we are finding the average squared residual between the original dataset of clean shot records and our prediction for these clean shot records based on our network's performance on removing the coherent noise in the noisy shot records in the time and frequency domains. In addition, we ensured that all pixel values whose magnitude is under the value of machine epsilon (approximately $10^{-15}$) were set equal to $0$ to ensure that there are no errors due to rounding performed by the model. Finally, we used the Adam optimizer as it has been a prefered optimizer for convolutional neural networks in the past \cite{check}. 
\subsection{Filtering Techniques}
In order to increase the accuracy of our model in the frequency domain, we utilize a low-pass Butterworth filter \cite{butterworth}, which smoothly removes all frequency information above a specified frequency value. Since there is no information in the input data in frequencies above 100 Hz, we set the cutoff frequency to be 100 Hz for the Butterworth filter so no additional information is added in higher frequencies during training.
\par
A summary of the project pipeline is shown in Figure \ref{fig:pipeline} below.

 \begin{figure}[htbp]
  \centering
  \includegraphics[width=\textwidth]{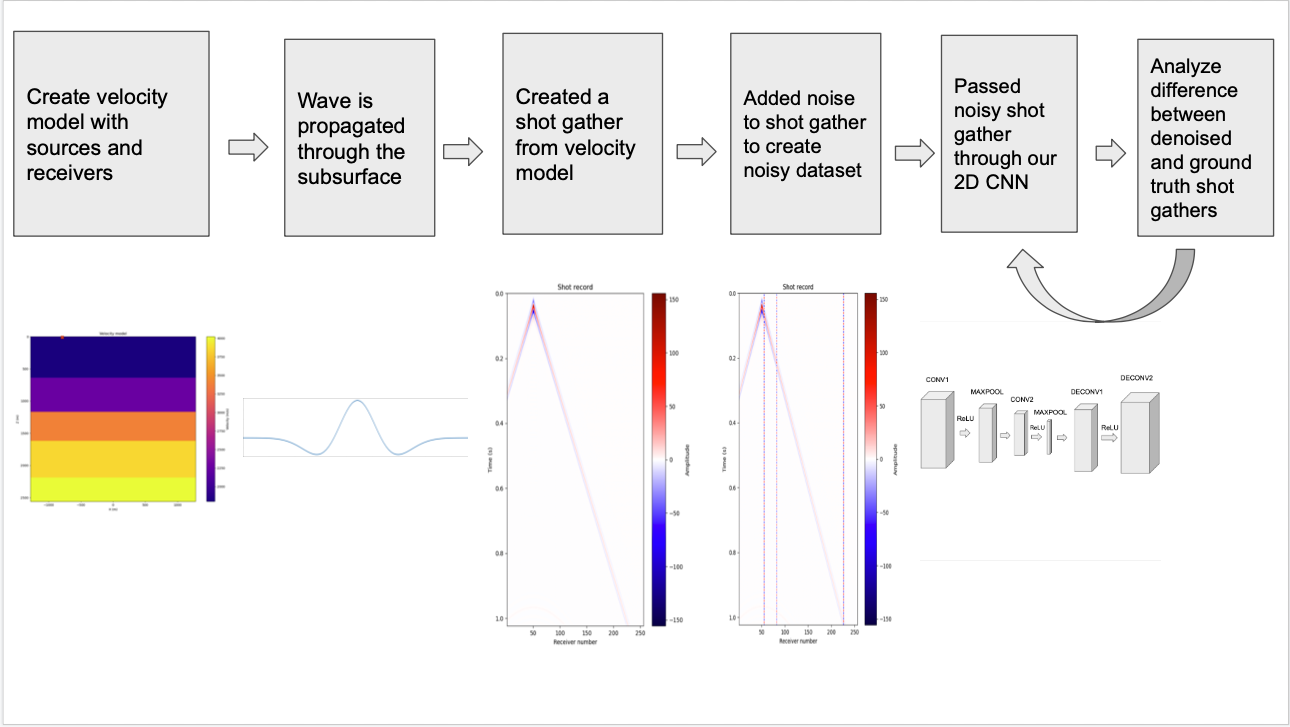}
  \caption{The figure above depicts the full project pipeline. The velocity model is created with sources and receivers on the surface, a wave modeled by the Ricker wavelet is propagated through the subsurface, a clean and noisy shot gather are created from the wave propagation, and the noisy shot gather is passed through the 2D CNN which removes the electrical noise.} 
  \label{fig:pipeline}
\end{figure}
\section{Main Results}
\begin{figure}[htbp]
  \centering
  \includegraphics[width=0.8\textwidth]{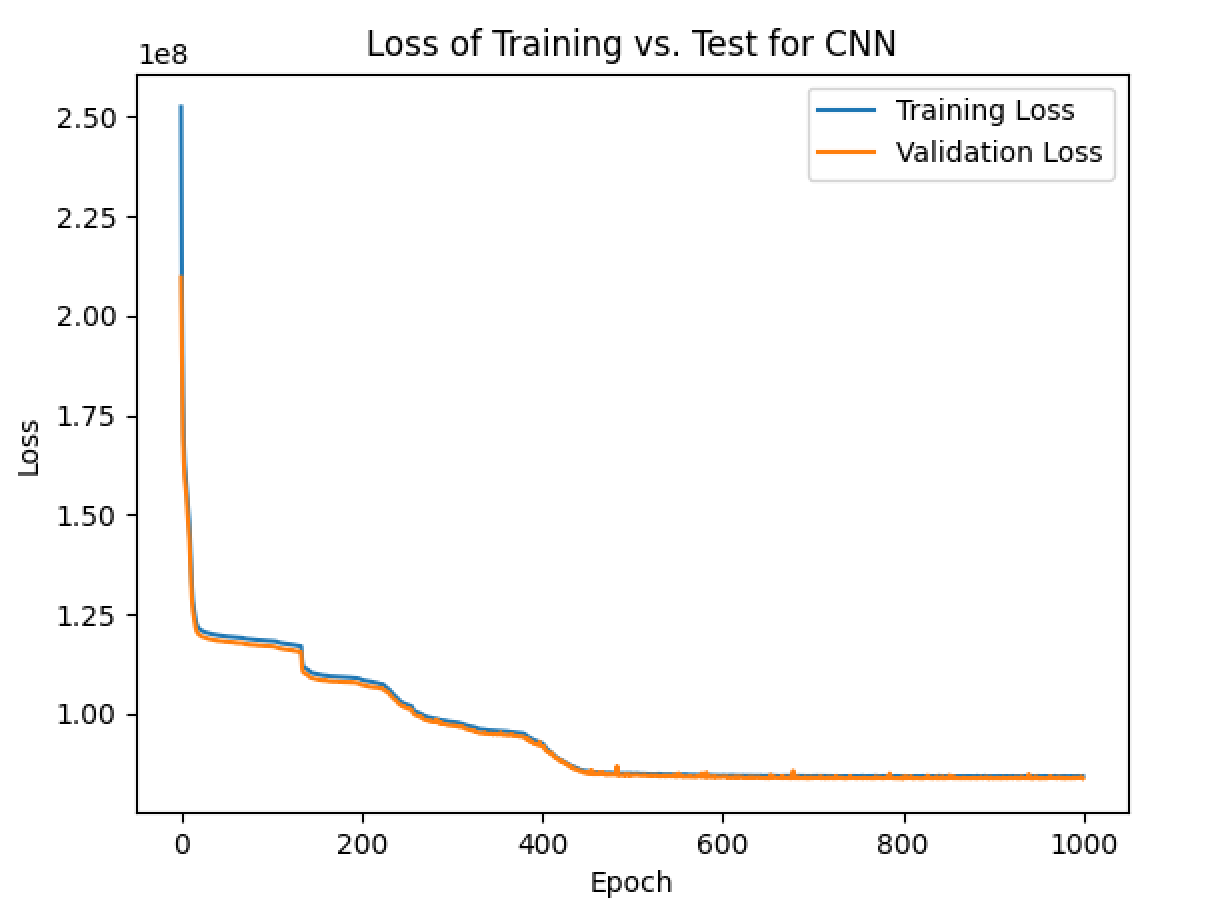}
  \caption{Plot of loss for training data and test data, where $80\%$ of the $10,000$ datasets were part of the training dataset and $20\%$ were part of the test dataset.}
  \label{fig:loss}
\end{figure}

To test the validity of our network, we have plotted graphs of training loss and validation loss, which are shown in Figure \ref{fig:loss}. We split our dataset of $10,000$ total noisy shot records from the NoisyShot dataset into $8000$ shot records ($80\%$) for the training data and $2000$ shot records ($20\%$) for the validation data. 
\par

\begin{figure}[htbp]
  \centering
  \includegraphics[width=\textwidth]{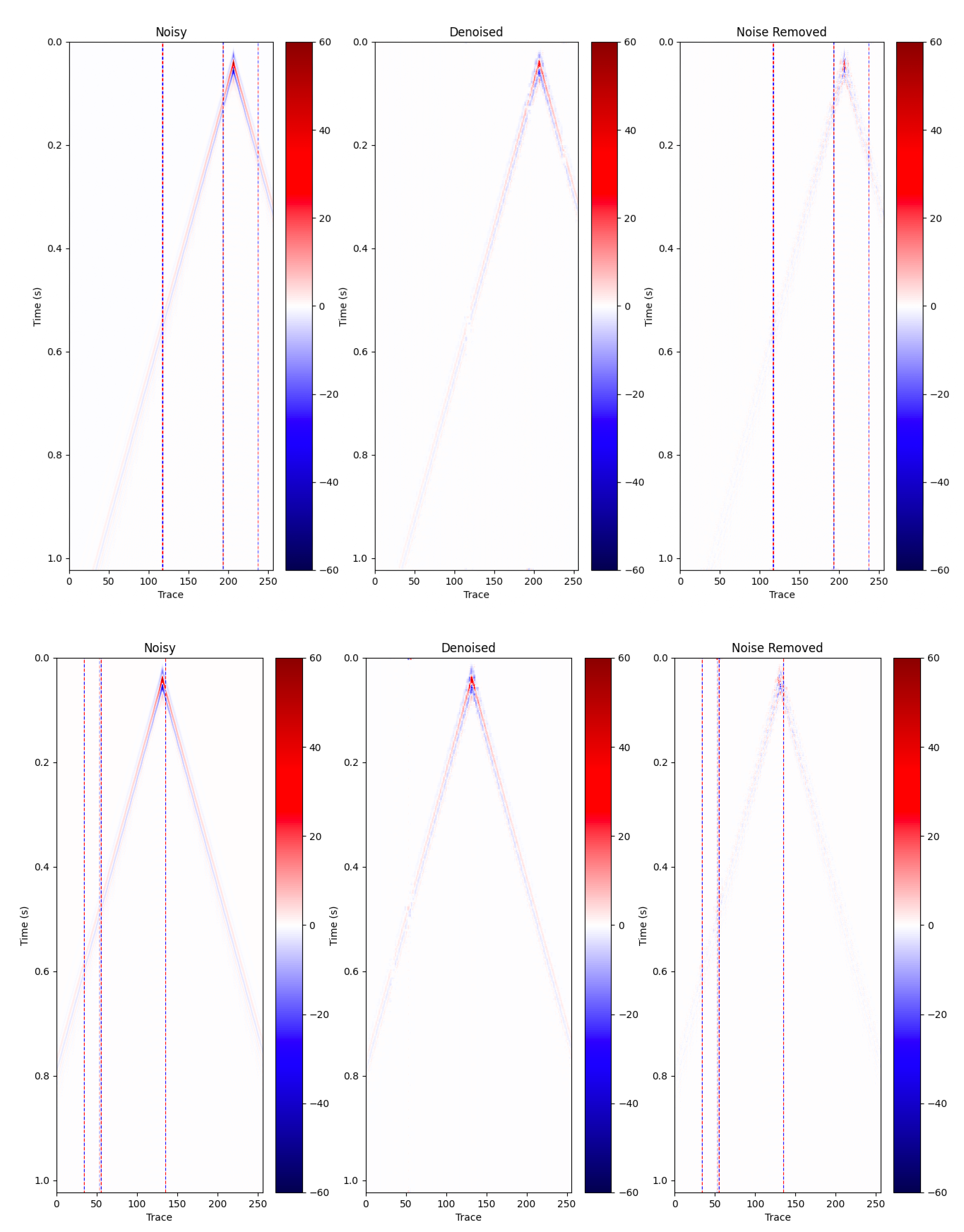}
  \caption{Two example datasets and their respective denoising results are shown (top and bottom rows). The original input noisy data (left), the denoised data (center), and the difference between the two, which is the removed noise (right) show the overall performance of the model is good at specifically identifying coherent seismic noise. However, a small portion of the shot record information is also removed.} 
  \label{fig:ayfig}
\end{figure}

We also examine the images before and after our denoising process. In Figure \ref{fig:ayfig}, we can see that the bottom shot record has coherent noise close to the peak of the seismic wave, which is one of the hardest scenarios in terms of removing coherent noise. However, the model still performs quite well, removing the coherent electrical noise while still retaining most of the important characteristics of the original shot record.
\begin{figure}[htbp]
  \centering
  \includegraphics[width=0.8\textwidth]{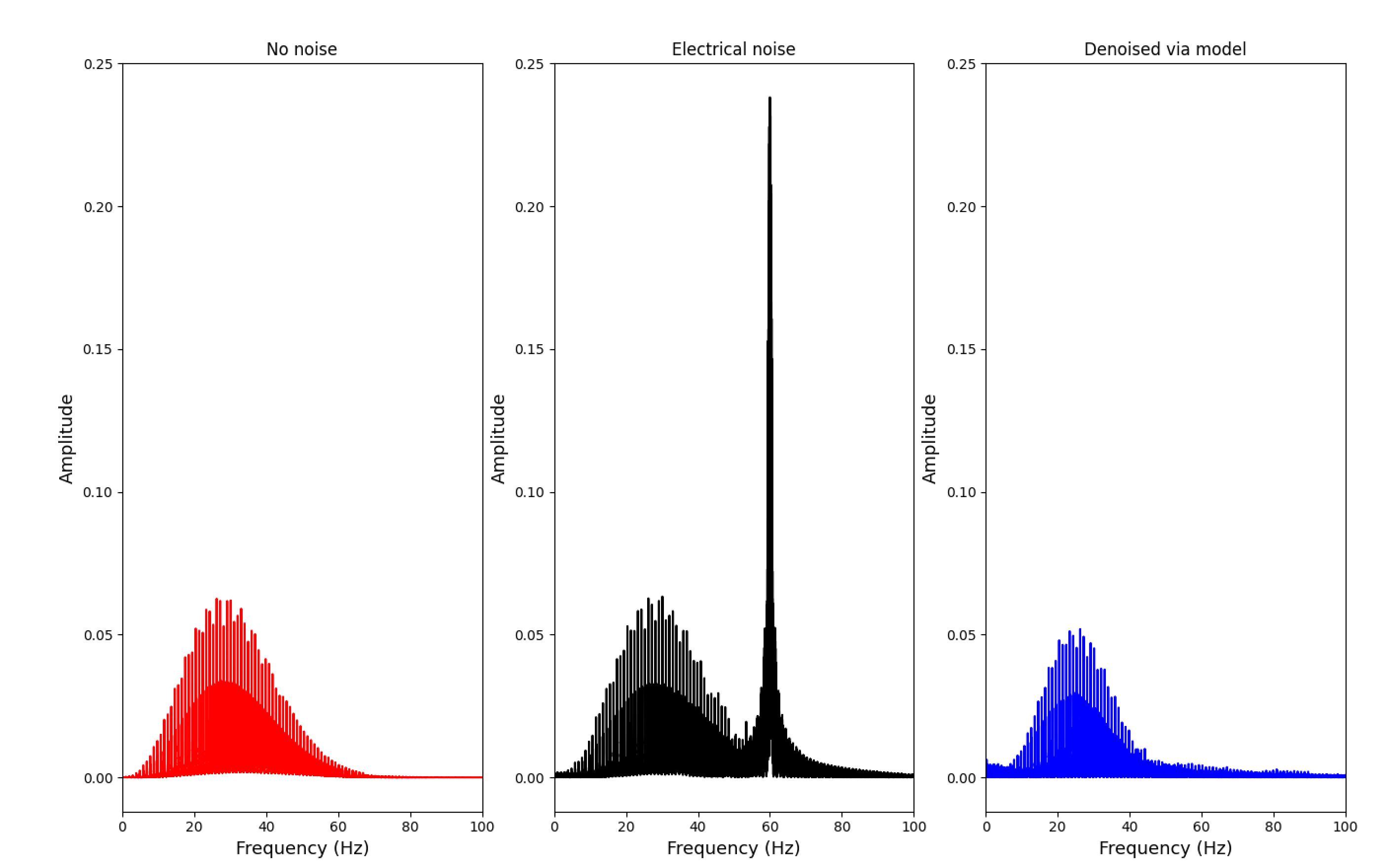}
  \caption{Frequency spectra of model performance. The left graph is the frequency spectra of the original shot record, the middle graph is the frequency spectra of the shot record with added electrical noise, and the right graph is the frequency spectra of the shot record denoised by our model.}
  \label{fig:fft}
  \end{figure}

\section{Analysis}

Our model removes coherent noise while retaining most of the characteristics of the original shot record, as shown in Figures \ref{fig:ayfig} and \ref{fig:fft}. Note that even when the coherent noise source is located near the shot location, the model still removes the noise quite well. The images depicting the differences between the truth shot record (without coherent noise) and the predicted shot record from our model are mostly correct, with small differences near the top of the shot record. In fact, even when the coherent noise corrupting the top of the shot record near the source location, the difference in the same shot record without the noise and the predicted shot record from our model was small. 
\par
Our graph of training loss versus testing loss in Figure \ref{fig:loss} demonstrates that neither overfitting nor underfitting occurred during our training process, as both the training and testing loss functions are decreasing at similar rates (for the most part) throughout the $1000$ epochs. We were able to achieve a normalized average L1 loss of approximately $2.3798 * 10^{-4}$ and a normalized average MSE loss of approximately $7.6384 * 10^{-6}$. Given the large scale of our data ($1024 \times 256$) and that a significant portion of our data consists of pixel values that are very large (in magnitude), this loss is low enough for us to conclude that the model efficiently removed the majority of the coherent noise while still retaining important features of the shot record.
\par
Finally, our graph of model performance in the frequency domain (as shown in Figure \ref{fig:fft}) demonstrates how well our model did as the entire spike of noise shown at 60 Hz was removed by our model. After applying the low-pass Butterworth filter upon this denoised shot record, we were able to retain most of the characteristics of the original shot record in the frequency domain while also completely removing the 60 Hz electrical noise.
\section{Further Works}
Though our model performs quite well, it can be improved to perform on different types of velocity models such as those with faults, folds, or salt. Another promising idea is to use sparse dictionary learning, which may allow for further interpretation of the underlying processes that separate the signal from the coherent noise. Finally, we would like to explore using adaptive coefficients in the model's loss function to prioritize either the frequency or time domains for optimal performance.

\section{Acknowledgments} 
The authors would like to acknowledge the MIT PRIMES-USA program for allowing them to conduct this research. The second author acknowledges support from the United States Department of Energy through the Computational Science Graduate Fellowship (DOE CSGF) under grant number DE-SC0019323.
\newpage

\bibliographystyle{siamplain}
\bibliography{references}

\begin{thebibliography}{1}

\bibitem{subtract}
{\sc K.~E. Butler and R.~D. Russell}, {\em Subtraction of powerline harmonics
  from geophysical records}, {GEOPHYSICS}, 58 (1993), pp.~898--903,
  \url{https://doi.org/10.1190/1.1443474},
  \url{https://doi.org/10.1190/1.1443474}.

\bibitem{butterworth}
{\sc S.~Butterworth}, {\em On the theory of filter amplifiers}, Experimental
  Wireless and the Wireless Engineer, 7 (1930).

\bibitem{arithmetic}
{\sc V.~Dumoulin and F.~Visin}, {\em A guide to convolution arithmetic for deep
  learning}, 2016, \url{https://arxiv.org/abs/arXiv:1603.07285}.

\bibitem{math}
{\sc J.~Gu, Z.~Wang, J.~Kuen, L.~Ma, A.~Shahroudy, B.~Shuai, T.~Liu, X.~Wang,
  G.~Wang, J.~Cai, and T.~Chen}, {\em Recent advances in convolutional neural
  networks}, Pattern Recognition, 77 (2018), pp.~354--377,
  \url{https://doi.org/10.1016/j.patcog.2017.10.013},
  \url{https://doi.org/10.1016/j.patcog.2017.10.013}.

\bibitem{mean}
{\sc H.~Karsl{\i} and D.~Dondurur}, {\em A mean-based filter to remove power
  line harmonic noise from seismic reflection data}, Journal of Applied
  Geophysics, 153 (2018), pp.~90--99,
  \url{https://doi.org/10.1016/j.jappgeo.2018.04.014},
  \url{https://doi.org/10.1016/j.jappgeo.2018.04.014}.

\bibitem{check}
{\sc D.~P. Kingma and J.~Ba}, {\em Adam: A method for stochastic optimization},
  2014, \url{https://arxiv.org/abs/arXiv:1412.6980}.

\bibitem{zscore}
{\sc S.~G. PATRO and D.-K.~K. Sahu}, {\em Normalization: A preprocessing
  stage}, IARJSET,  (2015), \url{https://doi.org/10.17148/IARJSET.2015.2305}.

\bibitem{rpca}
{\sc H.~Wang, H.~Zhang, and Y.~Chen}, {\em Sinusoidal seismic noise suppression
  using randomized principal component analysis}, {IEEE} Access, 8 (2020),
  pp.~152131--152145, \url{https://doi.org/10.1109/access.2020.3017690},
  \url{https://doi.org/10.1109/access.2020.3017690}.

\bibitem{hum}
{\sc J.~Xia and R.~D. Miller}, {\em Design of a hum filter for suppressing
  power-line noise in seismic data}, Journal of Environmental and Engineering
  Geophysics, 5 (2000), pp.~31--38, \url{https://doi.org/10.4133/jeeg5.2.31},
  \url{https://doi.org/10.4133/jeeg5.2.31}.

\end{thebibliography}

\end{document}